\documentclass[manuscript, screen, nonacm]{acmart} 
\AtBeginDocument{%
  }

\acmYear{2026}
\acmDOI{XXXXXXX.XXXXXXX}
\acmConference[CHI '26]{ACM CHI 2026 Workshop on
Human-Agent Collaboration}{April 13,
  2026}{Barcelona, Spain}
\acmISBN{978-1-4503-XXXX-X/2018/06}

\usepackage{xspace}
\newcommand{\method}{CompanionCast\xspace}

\begin{document}

\title{\method: Toward Social Collaboration with Multi-Agent Systems in Shared Experiences}

\author{Yiyang Wang}
\email{ywang3420@gatech.edu}
\affiliation{%
  \institution{Georgia Institute of Technology}
  \city{Atlanta}
  \state{Georgia}
  \country{USA}
}

\author{Chen Chen}
\author{Tica Lin}
\author{Vishnu Raj}
\author{Josh Kimball}
\affiliation{%
  \institution{Dolby Laboratories, Inc.}
  \city{Atlanta}
  \state{Georgia}
  \country{USA}
}

\author{Alex Cabral}
\author{Josiah Hester}
\affiliation{%
  \institution{Georgia Institute of Technology}
  \city{Atlanta}
  \state{Georgia}
  \country{USA}
}

\renewcommand{\shortauthors}{Wang et al.}

\begin{abstract}
Shared experiences are fundamental to social connection, yet media consumption is increasingly solitary. While AI companions offer real-time reactions and emotional regulation, existing systems either rely on single-agent designs or lack the social awareness and multi-party interaction required to replicate authentic group dynamics. 
We present \method, a general framework for orchestrating multiple specialized AI agents as social collaborators within a live shared context. 
\method integrates multimodal event detection, rolling context caching for improved grounding, and spatial audio to enhance co-presence. 
We validate \method through sports viewing, a domain with rich dynamics and strong social traditions. Pilot studies with soccer fans demonstrate that \method significantly improves perceived social presence and emotional sharing compared to solitary viewing. 
We conclude by discussing implications and open challenges for multi-agent systems as social collaborators in shared experiences. 
\end{abstract}


\begin{teaserfigure}
  \includegraphics[width=\textwidth]{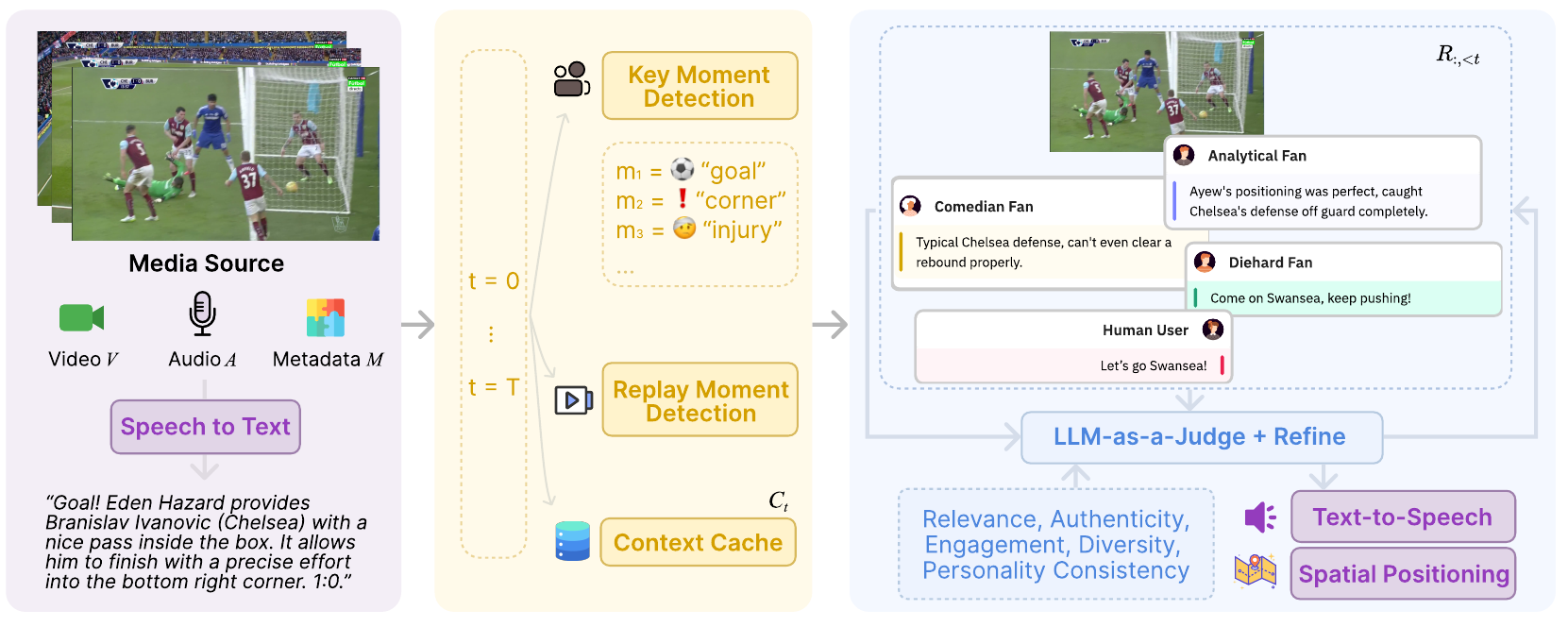}
  \caption{Overview of the system workflow. Media inputs are processed to extract captions. From these captions, the system identifies key moments and replay events that trigger agent interactions. Rolling context such as captions from the past one minute is cached and provided to the agents during dialogue generation. An LLM-as-a-judge module evaluates and refines the agent conversations. The finalized text is then converted to speech, after which spatial positioning is applied when producing the audio output.}
  \Description{Overview of the system workflow.}
  \label{fig:teaser}
\end{teaserfigure}

\received{15 February 2026}
\received[accepted]{3 March 2026}

\settopmatter{printacmref=false}
\setcopyright{none}
\renewcommand\footnotetextcopyrightpermission[1]{}
\pagestyle{plain}
\maketitle

\vspace{-3mm}
\section{Introduction}
\label{sec:intro}
\vspace{-1mm}

Shared experiences, such as co-viewing and collaborative learning, are fundamental to social connection and engagement~\cite{holt2024social, nowak_hear_2023}. 
Beyond consuming content, people share reactions, regulate emotions, and coordinate attention, making the experiences feel meaningful and connected. 
As activity participation increasingly moves to online platforms, people often lack social partners who can react in real time
and provide context-aware coordination during the experience~\cite{holt2024social, mukherjee_information_2017}.  
Recent advances in large language models (LLMs) and agentic systems \cite{li_camel_2023, crawford_bmw_2024, shu_towards_2024}
have inspired AI companions that react to unfolding context and adopt explicit social roles~\cite{wang2026mascot}. However, most  systems rely on single agent designs~\cite{kim_bleacherbot_2025,andrews_aicommentator_2024} or act as collaborative social systems with limited social awareness~\cite{li_big5-chat_2024, zhao2024competeai, ryu_cinema_2025}. Thus, these systems face
difficulty approximating real group dynamics, as they lack support for multiple concurrent collaboration needs (e.g., sensemaking, emotional expression, grounding/common ground, pacing, and inclusion). 

In this work 
we present \method (Fig.~\ref{fig:teaser}), a generalizable multi-agent framework to create shared experiences in social collaboration. 
\method is built around a live shared context and orchestrates role-specialized agents with spatial audio positioning 
using (i) event and moment detection, (ii) rolling context caching to support grounding, (iii) multi-agent dialogue generation, and (iv) evaluator-driven refinement to manage interaction quality. 
We evaluate \method in sports co-viewing, a high-dynamics setting with strong social norms, rapid context shifts, emotionally salient moments, 
and multimodal data (e.g. video, captions, and commentary).
This provides a natural testbed for stimulating human-agent social collaboration under time pressure and partial shared awareness. 
We conduct a pilot study with two participants, finding that \method increases the sense of shared experience.
We demonstrate feasibility and surface practical design and engineering insights, along with open challenges around coordination, transparency, and safeguards for future human–agent collaboration systems.

\vspace{-2mm}
\section{\method: Towards Shared Experiences in Social Collaboration}
\vspace{-1mm}
\method is a generalizable LLM-powered multi-agent framework that orchestrates AI companions $\mathcal{A} = \{a_1, a_2, \dots, a_n\}$  around multimodal video content for shared experiences. 
It supports mixed-initiative interactions where agents proactively generate responses $R$ based on salient and replay moments $E \subset \mathcal{S}$ and user inputs. 
We define the core logic through four functional components: 

\noindent \textbf{Multimodal Content Processing.} 
The system ingests a multimodal stream $\mathcal{S} = \{V, A, M\}$,  including video frames $V$, audio $A$, and metadata $M$, such as event descriptions. 
We use the SoccerNet dataset~\cite{giancola_soccernet_2018} and its Dense Video Captioning subset~\cite{Mkhallati2023SoccerNetCaption-arxiv} to access temporally-aligned caption data. 
A structured temporal context $\mathcal{C}_t = \{ s_\tau \in \mathcal{S} \mid t - \omega \leq \tau \leq t \}$, is maintained and shared across all agents. 
Here, $\omega$ is the look-back window. For sports co-viewing, we set $\omega=60$ seconds. 

\noindent \textbf{Multi-Agent Orchestration} 
Each agent $a_i \in \mathcal{A}$ is configured with a social roles (e.g. supporter, analyst, observer, or humorist) and maintains a personality profile $p_i$, domain knowledge $k_i$, and interaction style $s_i$. 
We instantiate three specialized fan agents with distinct personalities: (1) \textit{Die-Hard Fan}: an enthusiastic supporter of the user's chosen team, characterized by emotional expressiveness and celebratory language, (2) \textit{Analyst Fan}: a tactical analyst of the user's team, providing objective technical observations and performance commentary, and (3) \textit{Comedian Fan}: a sarcastic fan supporting the opposing team, introducing playful antagonism and humor to create conversational tension. 
The response $r_{i,t}$ for agent $a_i$ at time $t$ is generated conditioned on the user profile and shared context: $r_{i,t} \sim f_{\phi}(\cdot \mid p_i, k_i, \mathcal{C}_t, \mathcal{R}_{:,<t})$, where $f_{\phi}(\cdot | \cdot)$ is the LLM parameterized by $\phi$, $\mathcal{R}_{:,<t}$ is the response history of all agents. 
The agent orchestrator employs a triggering function to manage the transition from passive observation to active collaboration based on detected important moments, scene changes, or user interactions.

\noindent \textbf{Spatial Audio Rendering} 
To enhance co-presence, CompanionCast assigns each agent $a_i$ a distinct voice $v_i$ and stereo pan value $l_i$, decoding synthesized speech into a shared Web \texttt{AudioContext} and routing each audio buffer through a \texttt{StereoPannerNode} to render left–right spatial separation.
Agent speech is synthesized using ElevenLabs text-to-speech~\cite{noauthor_free_2025} with three distinct voice profiles matched to agent personalities.
These voice and spatial parameters are configurable to adapt to diverse content types and user preferences. 
During agent conversations, original match audio is automatically muted to ensure intelligibility. 

\noindent \textbf{Evaluator Agent} 
An evaluator meta-agent $\hat{a}$ assesses conversations across $5$ dimensions: relevance, authenticity, 
personality consistency, diversity, and engagement ~\cite{deriu2021survey, see2019good}. 
The evaluator provides both quantitative scores on a $[0, 10]$ Likert scale and qualitative feedback to enable iterative refinement.
This feedback loop runs during natural pauses in live content or asynchronously for recorded material, enabling a novel application of AI-in-the-loop quality control for real-time multi-agent systems.

The framework provides abstractions for content analysis, agent configuration, conversation orchestration, and quality evaluation, making it adaptable to various video viewing domains including sports, movies, documentaries, educational content, and entertainment shows. 

\vspace{-2mm}
\subsection{Interaction Implementation Details}
\vspace{-1mm}



For event detection, we employ two complementary annotation streams to cover key viewing moments: (1) important moments (e.g. goals, fouls, corners, penalties) identified via labels from the SoccerNet Dense Video Captioning dataset \cite{Mkhallati2023SoccerNetCaption-arxiv}, and (2) replay segments detected using temporal boundaries from the SoccerNet Replay Grounding dataset \cite{Deliège2020SoccerNetv2}. 

We leverage AutoGen~\cite{wu2024autogen} to instantiate multi-agent interactions. 
For different game scenarios (goals, corners, penalties, replay), we provide scenario-specific system prompts defining expected emotional intensities, interaction patterns, and conversation dynamics.
The fan agents use Claude Sonnet 4~\cite{claude} with a temperature $\tau$ of $0.7$ to encourage conversational diversity while maintaining coherence, while the evaluator $\hat{a}$ uses OpenAI GPT-4o~\cite{gpt4o} with $\tau=0.2$ to balance consistency with diversity of feedback.

The system executes a multi-round conversation protocol: the agent team generates initial responses, received evaluator feedback, and performed iterative refinements before presenting the final conversation to users. During important match moments, we execute $k=3$ refinement rounds, while replay moments and user queries utilize $k=1$ and $2$ rounds respectively to balance quality with timing. 

The system proactively initiates multi-agent conversations at detected important and replay segments. To prevent conversational overlap and maintain viewing flow, conversations are subject to a minimum 30-second separation constraint (or 15 seconds for high-intensity periods). User-initiated conversations are supported at any time via voice or text, with a maximum of 3 messages per initial agent reaction round for conversational brevity.


\vspace{-1mm}
\section{Results and Discussion}
\vspace{-1mm}
\noindent \textbf{Experiment Setup}
\begin{figure}[htbp]
  \centering
  \vspace{-2mm}
  \begin{minipage}[t]{0.45\linewidth}
    \centering
    \includegraphics[width=\linewidth]{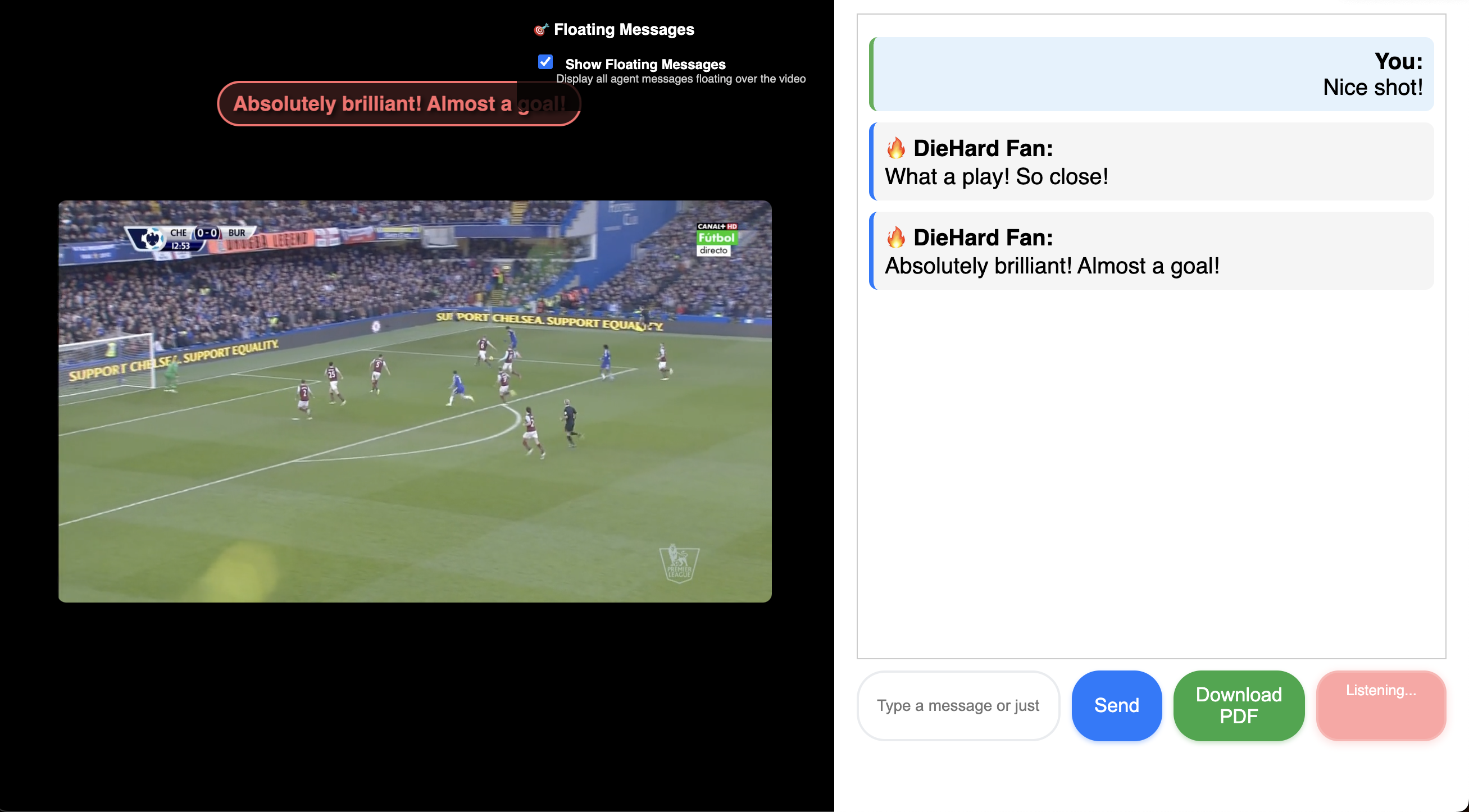}
    \vspace{-2mm}
    \caption{Implemented system used in the pilot user study.}
    \label{fig:actual_system}
  \end{minipage}\hfill
  \begin{minipage}[t]{0.49\linewidth}
    \centering
    \includegraphics[width=\linewidth]{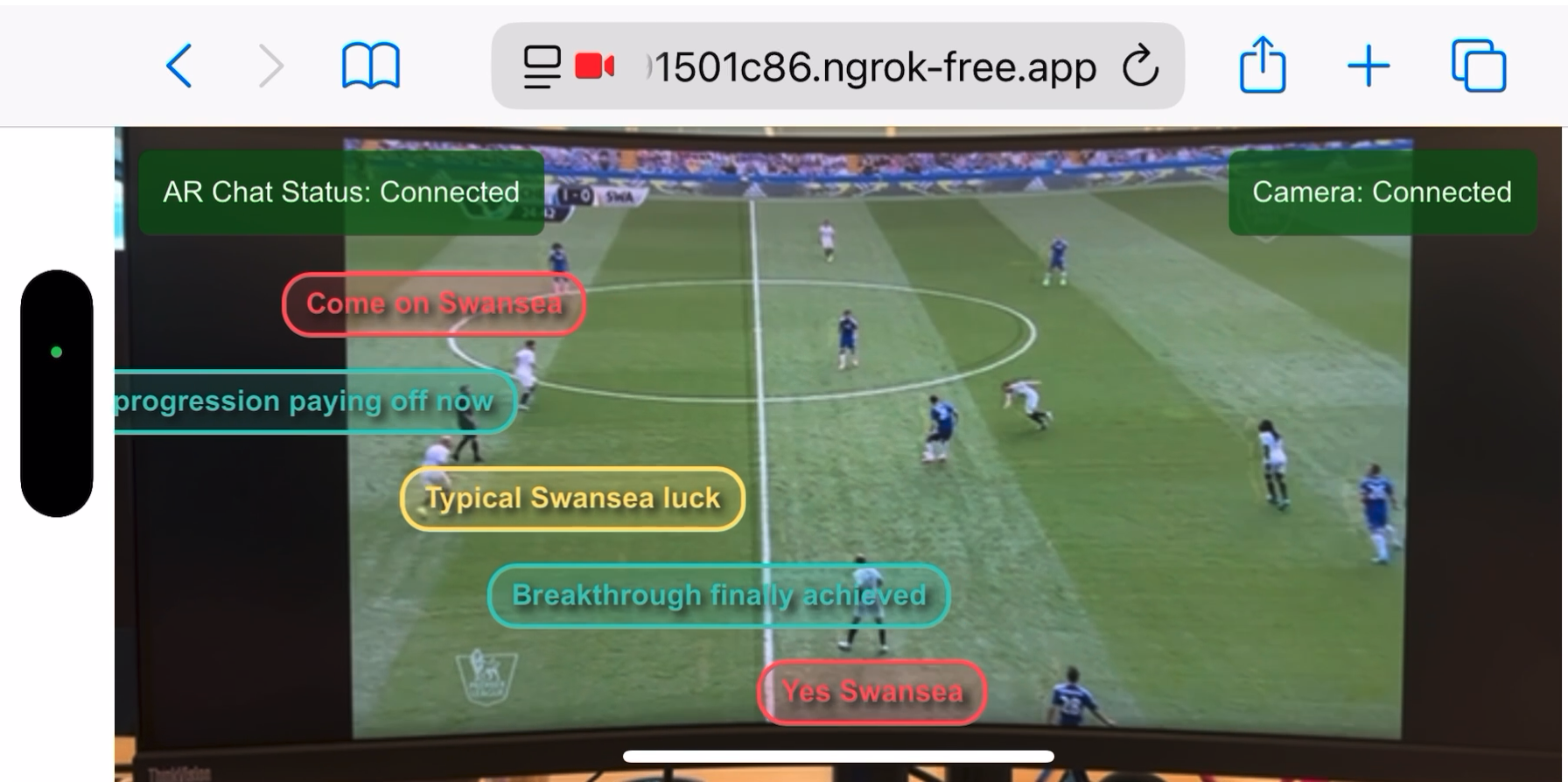}
    \vspace{-4mm}
    \caption{Exploratory AR prototype built with WebAR and demonstrated on a mobile phone.}
    \label{fig:ar}
  \end{minipage}
  \vspace{-4mm}
\end{figure}
To validate \method, we implement a soccer viewing web application (Fig.~\ref{fig:actual_system}) and conduct an in-person within-subjects study. Participants watched two $\sim$5-minute clips
containing salient events (goals, fouls, replays) under two counterbalanced conditions: 1) baseline, watching original clips alone, and 2) \method, viewing with AI companions. For \method, participants engaged in at least one user-initiated interaction. 
We recruited two adult soccer fans (both Asian males, age 28) with prior experience watching soccer socially. After viewing, participants completed a Likert-scale questionnaire assessing perceived agent performance, user engagement \cite{see2019good}, social co-presence \cite{nowak_hear_2023}, and conversation quality, followed by a semi-structured interview. All data were anonymized.

We identify four primary themes from the pilot study: (1) \textbf{Multi-Agent Presence Supported a Sense of Shared Experience}, (2) \textbf{Multimodal Agents Were Distinguishable and Socially Interpretable}, (3) \textbf{Contextually Grounding in Shared Experience}, and (4) \textbf{Multi-Agent Interaction Prompted Social Reflection}.

Both participants indicated moderate increases in perceived social co-presence (both rated 4/5) and immersion (both rated 3/5), and high willingness to share emotions with CompanionCast (P1 rated 4/5 and P2 rated 5/5). Behavioral engagement metrics supported this, with P1 initiating two messages during the session and P2 initiating four. Both participants described the experience as immersive, particularly when agents proactively reacted during exciting moments. Agent reactions helped sustain attention and reduced the likelihood of missing key events, suggesting that event-based triggering and shared context access effectively grounded agent interaction in the shared experience. These findings suggest that the presence of multiple agents enhanced the shared viewing experience and contributed to a greater sense of watching "with others".

Both participants also found the agents to be distinguishable and socially interpretable, and appreciated the diversity of perspectives across agents. Multimodal cues, including distinct voice identities and spatialized audio, further supported agent differentiation, even when effects were subtle. Furthermore, ratings for relevance and authenticity were moderate to high, and participants frequently interpreted agent behavior through a social lens, comparing agents to human co-viewers such as fellow fans. These findings suggest that CompanionCast supported engagement with agents as distinct social collaborators rather than as a single unified voice or background commentary.

Taken together, these findings suggest that multi-agent systems can function as social collaborators in shared experiences when coordination, role differentiation, contextual grounding, and interaction modality are treated as key design considerations. The effectiveness of event-driven reactions highlights the importance of temporal alignment and proactive participation in collaborative agent behavior, while participants' appreciation of role diversity underscores the value of distributing social functions across agents rather than concentrating them in a single persona. Finally, the role of multimodal cues in supporting interpretability suggests that collaboration quality depends not only on what agents say, but on how their actions and identities are presented to users. 

\vspace{-1mm}
\section{Conclusion and Future Work}
\vspace{-1mm}
This work presents CompanionCast, a multi-agent framewor k that enables social collaboration in shared experiences. Future work can examine broader domains, larger studies, and adaptive and embodied interaction paradigms (Fig.~\ref{fig:ar}).



\bibliographystyle{ACM-Reference-Format}
\bibliography{references, other_refs}

\end{document}